\documentclass[aps,twocolumn,showpacs,english,prl,superscriptaddress]{revtex4-1}

\usepackage{graphicx}% Include figure files
\usepackage{dcolumn}% Align table columns on decimal point
\usepackage{bm}% bold math
\usepackage[T1]{fontenc}
\usepackage[latin9]{inputenc}
\usepackage{textcomp}
\usepackage{amsmath}
\usepackage{graphicx}
\usepackage{amssymb}
\usepackage{units}

\begin{document}

\title{Long range correlations in a 97\% excitonic one-dimensional
  polariton condensate}

\author{Aurélien Trichet}
\affiliation{Institut Néel, Université Joseph Fourier and CNRS, B.P. 166, 38042 Grenoble, France}
\affiliation{Department of Materials, University of Oxford, Parks Road, Oxford OX1 3PH, United Kingdom}
\author{Emilien Durupt}
\affiliation{Institut Néel, Université Joseph Fourier and CNRS, B.P. 166, 38042 Grenoble, France}
\author{François Médard}
\affiliation{Institut Néel, Université Joseph Fourier and CNRS, B.P. 166, 38042 Grenoble, France}
\affiliation{Institut Pascal (IP), Clermont Université, Université Blaise Pascal, BP 10448, F-63000 Clermont-Ferrand, France}
\author{Sanjoy Datta}
\affiliation{Université Grenoble I and CNRS, Laboratoire de Physique et Modélisation des Milieux Condensés,
UMR 5493, B.P. 166, 38042 Grenoble, France}
\author{Anna Minguzzi}
\affiliation{Université Grenoble I and CNRS, Laboratoire de Physique et Modélisation des Milieux Condensés,
UMR 5493, B.P. 166, 38042 Grenoble, France}
\author{Maxime Richard}
\email{maxime.richard@grenoble.cnrs.fr}
\affiliation{Institut Néel, Université Joseph Fourier and CNRS, B.P. 166, 38042 Grenoble, France}


\begin{abstract}
We report on the realization of an out-of-equilibrium polariton condensate under pulsed excitation in a one-dimensional geometry. We observe macroscopic occupation of a polaritonic mode with only $3\%$ photonic fraction, and a nature  strikingly close to that of a bare exciton condensate. With the help of this tiny photonic fraction, the condensate is found to display first-order coherence over distances as large as 10 microns. Based on a driven-dissipative mean field model, we find that the correlations length is limited by the effects of a shallow disorder under non-equilibrium conditions.
\end{abstract}

\date{\today}

\maketitle

By cooling down a gas of bosons below a critical temperature, a large fraction of the gas suddenly "condenses" into a single quantum state. This phase is characterized by the formation of off-diagonal long range order, that reaches infinity only in the specific case of a uniform
Bose-Einstein condensate \cite{Penrose1956}. The transition to the quantum regime occurs when the average correlation length $\lambda_c$ within the gas becomes comparable with the mean interparticle distances $d\simeq n^{-1/3}$, where $n$ is the particle density. This quantum-coherent regime is not restricted to Bose-Einstein condensates. In more complex quantum fluids, the range of the correlations is in general large but finite, and its decay is determined by the interactions strength, the dimensionality-dependent quantum fluctuations, the losses, and the disorder. For example, the first quantum many-body bosonic system to be identified as such was superfluid Helium IV. Indeed London has been the first in 1938 to relate its behavior with the formation of a macroscopic "condensate" \cite{London1938} in spite of its liquid phase, and strong interactions.

Later, the notion of quantum gases has been shown to be a pertinent one for excitonic gas in solid state environment. Indeed, Blatt and Moskalenko predicted that they could undergo Bose-Einstein condensation at elevated temperature ($T\simeq 1K$) thanks to their very light mass \cite{Blatt1962,Moskalenko1962}. Pioneering experiments have been carried out since then, showing that it is in fact difficult to reach the quantum regime with excitons. The most striking examples are exciton gases in cuprate semiconductor materials \cite{hulin1980,Snoke1987,Lin1993} and in coupled quantum wells \cite{Butov2001}. In these systems, excitons have a lifetime long enough to thermalize so that the correlation length is set by the thermal de Broglie wavelength $\lambda_x=\sqrt{2\pi \hbar^2/m_xk_BT}$, where $m_x$ is the exciton effective mass, $k_B$ is the Boltzmann constant and $T$ the temperature. $\lambda_x$ typically amounts to a few tens of nanometers at $T=1K$. The quantum regime is in principle achievable since at condensation threshold the interparticle distance is still larger than the Bohr radius. However, at such densities the long range nature of Coulomb interactions causes excitons to scatter strongly with each other, resulting in Auger driven heating of the gas, or biexciton formation \cite{Jang2006a}. Excitons can also get trapped locally due to short range disordered potential \cite{Butov1999}. All these mechanisms can be detrimental enough to prevent reaching quantum degeneracy. It is only very recently that a first indication of long-range correlations in a bare exciton gas has been reported in a ultra-high quality coupled quantum wells system \cite{High2012}. But a clear cut demonstration has remained elusive so far.

Over the last decade, another strategy has proven more fruitful to achieve quantum phase of Bose gases in solid state environment. In semiconductor microcavities in the strong coupling regime, elementary excitations are not bare excitons  but exciton-polaritons, which can be understood as excitons strongly dressed with the cavity photon field. With polariton gases, Bose-Einstein condensation \cite{Kasprzak2006} and superfluid behavior \cite{Amo2009a,Lagoudakis2009,Amo2011} have been demonstrated unambiguously together with many interesting new features resulting from their out-of-equilibrium situation. The reason for this success is that polaritons have a mass $m_p$ typically four orders of magnitude lighter than that of bare excitons $m_x$. Whether polaritons are at thermal equilibrium or not, this ultralight mass is the key feature. Indeed, in the latter case usually referred as polariton lasing \cite{Kasprzak2008}, the maximum kinetic energy of the condensate is limited by the finite polariton linewidth $\hbar\Gamma$ instead of $kT$ in the thermal case, i.e. $\hbar^2k_p^2/2m_p=\hbar\Gamma$ where $\lambda_p=2 \pi \hbar/k_p$ is the correlation length. Therefore, with a correlation length typically 100 times larger than bare excitons, polaritons enter the quantum regime at a density six orders of magnitude smaller. At such small density, Coulomb  interactions between polaritons \cite{Vladimirova2010} have a much less detrimental effect for the formation of long range correlations.

Interestingly, the polariton effective mass $m_p$ depends on the photonic weight $|G|^2$ of the dressed state according to the simple relation:
\begin{equation}
\frac{1}{m_p}=\frac{|G|^2}{m_g}+\frac{|X|^2}{m_x}
\label{mp}
\end{equation}
where $m_x$ is the undressed exciton mass, $m_g$ is the bare cavity effective mass, $|X|^2=1-|G|^2$ is the excitonic weight of the dressed state. This simple equation has a striking consequence: since typically $m_g/m_x<10^{-4}$, even excitons with a few percent of photonic dressing might become light enough to overcome the difficulties encountered with gases of bare excitons.

In this letter, we show that in ZnO microwires featuring one dimensional confinement, a polariton condensate with only $|G|^2=3\%$ photonic fraction is obtained above a threshold density by polariton lasing mechanism. We find that in spite of this largely dominant excitonic nature, the measured correlation length of this nearly excitonic condensate is surprisingly large, i.e. comparable with that previously reported in condensate of $\sim 50\%$ photonic polaritons. Using a mean field model, we discuss in detail the measured first order correlations function and find that its decay is well accounted for by the non-equilibrium features of the measurement in the presence of a shallow disorder and weak interactions.

\begin{figure}[t]
\includegraphics[width=0.9\columnwidth]{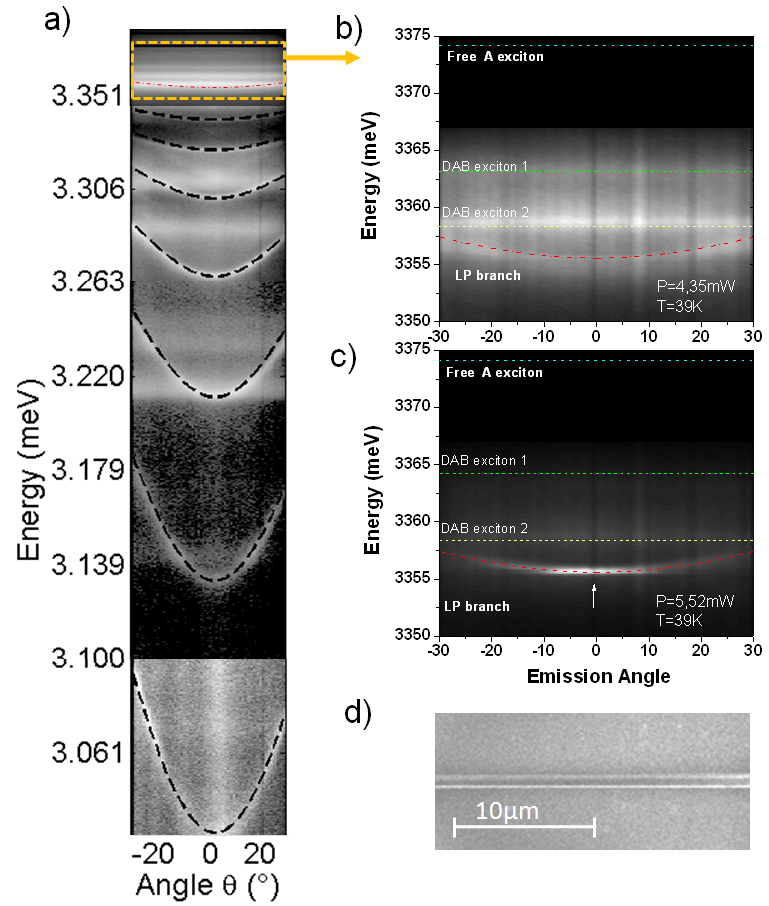}
\caption{(Color online) a) Measured photoluminescence in momentum space over a large spectral range. Several one dimensional lower polariton branches are visible, each corresponding to a different transverse state. Dashed lines : fit of the dispersion branches using a semiclassical model (see text). (b) Spectral zoom over the spectral range of interest, below the lasing threshold, and (c) above. (d) scanning electron micrograph of the microwire.}
\label{fig1}
\end{figure}

ZnO microwire polaritons result from the strong coupling regime between the whispering gallery modes (WGMs) and the bulk ZnO excitonic transitions (usually labeled A, B and C in order of increasing transition energy, their splitting resulting from the wurtzite uniaxial symmetry \cite{Klingshirn2010}). Due to momentum discretization in the direction perpendicular to the wire axis, exciton-polaritons are intrinsically one-dimensional in this system \cite{Trichet2011}. Very large Rabi splittings ranging from $200meV$ to $300meV$ have been reported already in such microwires \cite{Sun2008,Trichet2011} which result from the large intrinsic dipole moment of ZnO exciton, and from the large integral overlap between the WGMs and the excitonic medium. Moreover, owing to a very small excitonic Bohr radius of $a_B=2.3nm$, polaritons in ZnO are weakly interacting as compared to other materials.

One-dimensional polariton gases have been studied already in other systems and materials: long range correlations \cite{Manni2011} and complete measurement of the one-body density matrix \cite{Manni2012} have been reported in linear defects of a Telluride microcavity. Similar results have been obtained in high-Q Arsenide-based microcavities etched into linear waveguides \cite{Wertz2010}. In all these works the  photonic fraction is  $|G|^2\lesssim 50\%$. The realization of polariton states with a large excitonic fraction is an experimental challenge since it is difficult to obtain well-defined polariton mode of energy close to that of the bare exciton reservoir. To do so, one needs a very large Rabi splitting and a very low excitonic inhomogenous broadening, so that very excitonic modes remain in the motional narrowing regime \cite{Savona1997a}, i.e. the modes remain spectrally narrow and well separated from the exciton reservoir. ZnO microwires fulfill both requirements:  they display the largest Rabi splitting ever reported in inorganic semiconductor materials and an  excitonic homogeneous plus inhomogeneous broadening which we measured to be as low as $1.8meV$ at cryogenic temperature, this latter feature being due to the  excellent crystalline quality of the bulk excitonic medium.

Experiments on cold ($T=40K$) single crystalline ZnO microwires of hexagonal cross-section, with typically $50µm$ length and $2µm$ diameter are reported in this letter. A micro-photoluminescence setup is designed with spatial and angle-resolved detection capabilities. In addition a modified Michelson interferometer \cite{Kasprzak2006,Manni2011} is inserted in the detection path in order to perform spatio-temporal first order correlations measurements along the 1D polariton condensate. The polariton population is optically excited along a chosen segment of the wire (typically over $20 \mu m$), with a frequency doubled Titanium-Sapphire picosecond laser tuned at resonance with the A exciton plus $10meV$ excess energy. Considering the exciton binding energy of $60meV$, no free charges were excited in this way, nor C excitons.

In a first step, the Rabi splitting of the microwire is accurately determined using energy and angle-resolved photoluminescence measurements realized under weak excitations, as  shown in Fig.\ref{fig1}.a. In this plot, several lower polariton branches are visible, each corresponding to the coupling with a different WGM. Following a semiclassical approach \cite{Savona1999} detailed in \cite{Trichet2012}, the lower polariton dispersion branches are fitted with a single set of two free parameters, i.e. the overlap integral and the Rabi splitting $\hbar\Omega$. In this example, An overall Rabi splitting of $\hbar\Omega=288meV \pm 29meV $ is found where both A and B exciton contributions are included; C exciton contribution is vanishing due to polarization selection rule. This value hardly changes from wires to wires of diameter $d>1\mu m$.

\begin{figure}[t]
\includegraphics[width=\columnwidth]{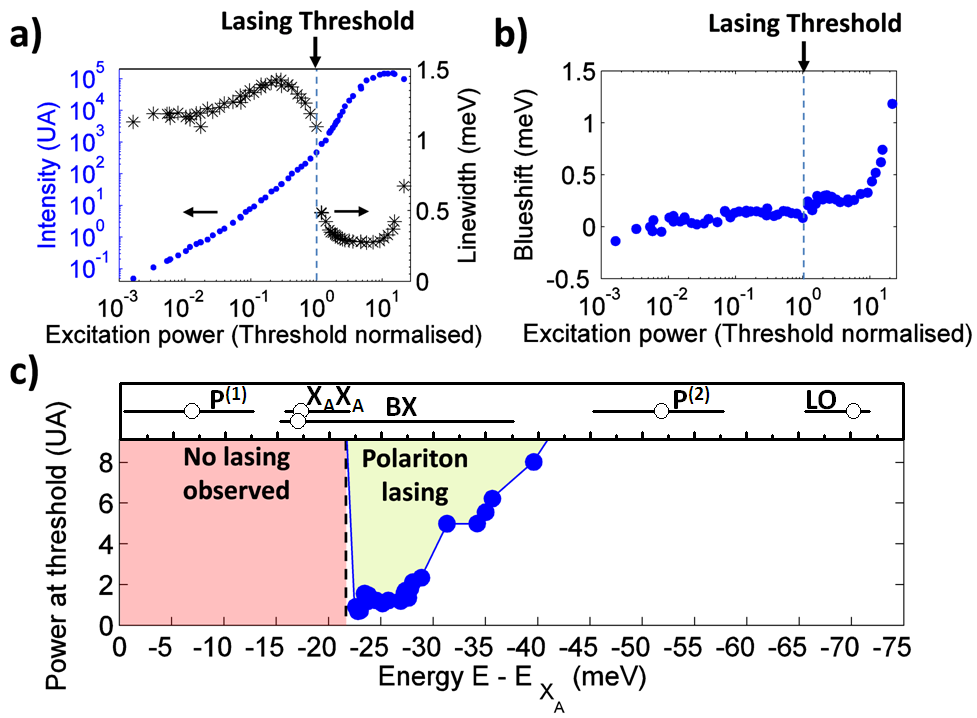}
\caption{(Color online) a) Output polariton emission intensity (circles) and linewidth of the emission (stars) vs input excitation. Vertical dashed line: polariton lasing threshold. b) Blueshift of the polariton emission over the same range of excitation power. c) Filled blue circles: measured excitation threshold for polariton lasing vs polariton branch energy. Hollow circles above: theoretical gain peaks and typical linewidth (horizontal black line) of the mechanisms considered in the text.}
\label{fig2}
\end{figure}

Upon increasing excitation density, we observe the sudden appearance of a macroscopic population of the $k_z=0$ polariton state in a polariton branch situated $20meV$ below the bare A exciton (cf. Fig.\ref{fig1}.b and Fig.\ref{fig1}.c). Due to the pulsed excitation, this effect is of the polariton laser type. The latter is identified by studying the behavior of the emission intensity $I$ as a function of the excitation power $P$. The results are shown in Fig.\ref{fig2}.a. At low density, the polariton emission is linear with the excitation intensity, as expected from phonon-assisted relaxation of reservoir excitons. At intermediate density, two-body scattering in the exciton reservoir starts to contribute to the dynamics and causes a nonlinearity with an approximate $I \sim P^{1.8}$ dependence. Scattering of the exciton bath with polaritons also causes a slight increase of the linewidth in this regime. Above threshold, in spite of the time integrated nature of the measurement, a  much stronger nonlinearity shows up accompanied by an abrupt narrowing of the linewidth, whose size is limited by the apparatus resolution of $200\mu eV$.

This is the usual behavior expected in the case of polariton lasing, with two unusual characteristics. First, unlike  microcavities made of other materials like Arsenides, Tellurides or Nitrides, the polariton emission hardly blueshifts at all at threshold until the excitation intensity becomes ten times larger. This is due to the very small excitonic Bohr radius in ZnO which results in lowered inter-polariton Coulomb interaction. It also guarantees \emph{a fortiori} that lasing occurs in the strong-coupling regime. Second, we find that the photonic fraction of the condensate is as low as $|G|^2=3\%$. We extract this value from the polariton laser energy $\hbar\omega_p$ and Rabi splitting, using the expression of the electromagnetic energy density in  dispersive media \cite{Philbin2011}:
\begin{equation}
\frac{1}{|G|^2}=1+\frac{\Omega^2}{4(\omega_p-\omega_x)^2},
\end{equation}
where the A and B excitonic transition energies $\hbar\omega_{xA}$ and $\hbar\omega_{xB}$ have been  merged into a single one since $\Omega \gg \omega_{xB}-\omega_{xA}$. Another estimate $|G|^2=2.4\%$ is consistent with the previous one and does not require the knowledge of $\Omega$. It is obtained from the measurement of the effective masses and using Eq.(\ref{mp}) which simplifies into $|G|^2\simeq m_g/m_p$. The polaritonic effective mass is deduced from the measured curvature of the dispersion relation in Fig.\ref{fig1}.b, and the bare photonic mass amounts to $m_g=n_0\hbar\omega_p/c^2$ with $n_0=2.35$, the background dielectric constant in ZnO and $c$ the speed of light.

Among the numerous available lower polariton modes, why does the system choose a mode for lasing with such a large excitonic fraction? In fact, as usually assumed with polariton gases, the polariton laser is fed by a stimulated inelastic scattering mechanism where free excitons in the reservoir are transferred towards the polariton states. In order to find out at which polariton energy the lowest threshold occurs, we measured the excitation threshold for lasing versus the polariton energy, and compared the energy of this lowest threshold with various possible scattering mechanisms. The results are shown in Fig.\ref{fig2}. It follows that due to the large Rabi splitting all possible mechanisms would provide laser gain in a spectral region where the polaritons have a large excitonic fraction. Moreover, according to our measurement (blue rectangle in Fig.\ref{fig2}.c), a gain mechanism based on biexciton ('$\text{X}_A\text{X}_A$' in Fig.\ref{fig2}.c) recombination is likely to be involved, as reported previously in Nitride planar microcavity \cite{Corfdir2012}. A mechanism based on scattering between a free A exciton and a bound exciton ('BX' in Fig.\ref{fig2}.c) could also be a good candidate \cite{Honerlage1976}. On the other hand, some other mechanisms, although popular in the domain of photon lasing in bulk ZnO slab, can be unambiguously ruled out by this comparison. These include exciton-LO phonon scattering ('LO' in Fig.\ref{fig2}.c), and free A exciton-exciton scattering resulting in the so-called P-band ($P^{(1)}$ and $P^{(2)}$ in Fig.\ref{fig2}.c) \cite{Klingshirn1981}.

\begin{figure}[t]
\includegraphics[width=\columnwidth]{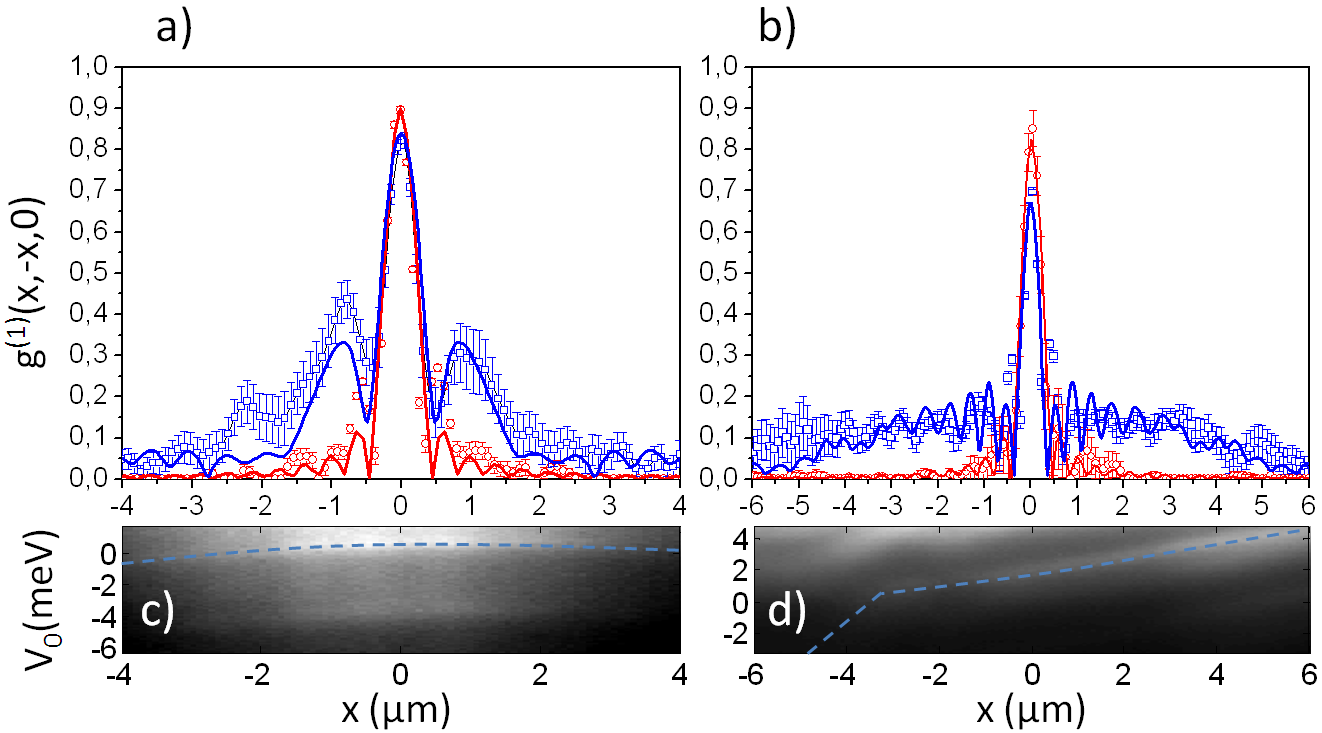}
\caption{(Color online) Measured $g^{(1)}(x,-x)$ functions on two points of two different wires, below (red circles) and above (blue squares) threshold. Condensate energies: $3351meV$ (a) and $3353.2meV$ (b). Solid lines: theoretical fit of the data: below (red line) and above (blue line) threshold (details in the supplemental material). c) and d) microphotoluminescence images obtained  below threshold. Dashed blue line: potential experienced by the polariton condensate of (a) and (b).}
\label{fig3}
\end{figure}

In order to determine the nature of this quasi-excitonic polariton condensate confined in a one-dimensional waveguide, we performed spatial correlations measurements at zero time delay. This is obtained by interferometric measurements performed with the condensate emission \cite{Kasprzak2006}. Since the excitation is pulsed, we measured the time-integrated normalized correlation function
\begin{equation}
g^{(1)}(x,-x)=\frac{\int_T \psi(x,t)\psi^\dagger(-x,t)dt}{\sqrt{\int_T |\psi(x,t)|^2dt \int_T |\psi(-x,t)|^2dt}}
\label{measured_g}
\end{equation}
where $\psi$ is the condensate wavefunction and the point $x=0$ is chosen in the middle of the polariton condensate, homogeneously excited over $20µm$ by an ellipsoidal laser spot. Since in an out-of-equilibrium situation two or three modes can sometime contribute to the condensate \cite{Krizhanovskii2009}, the obtained interferogram is spectrally resolved in order to select only the dominant transverse mode (The full $g^{(1)}(x,-x,\omega)$ spectra are shown in the supplemental information). The results are shown in Fig.\ref{fig3}.a and Fig.\ref{fig3}.b in two different wires of similar diameters for the sake of generality. Below threshold, the correlations are very short, far below our experimental resolution of $1\mu m$,  and entirely determined by the numerical aperture of our detection objective \footnote{See details in the Supplemental Material}.
Above threshold, in the lasing regime, we observe that a significant amount of correlations build up over $\sim 10 \mu m$ length. This measurement shows that in spite of a largely excitonic nature, the enhanced interactions and the much larger mass (by a factor of $\sim 15$) with respect to 50\% photonic polaritons do not prevent the formation of correlations over long distances. The measured $g^{(1)}(x,-x)$ also exhibits modulations and decay versus $x$ slightly different for both condensates. By analyzing in detail these features, we can determine the nature of the condensate, i.e. whether the correlations length is limited by the one dimensionality, the disorder, %or although weak,
or the interactions. Note that in our experiment, unlike in most previous works, the correlations length is not limited by the excitation spot size.

To understand the role of the disorder and interactions, the condensate wavefunction  $\psi(x,t)$ is simulated \cite{muru2009} by a mean-field generalized Gross-Pitaevskii equation including loss and gain, coupled via the gain term to a rate equation governing the dynamics of the excitonic reservoir density $n_R(x,t)$ \cite{Wouters2007c}, which in the case of a delta-pulsed pump can be solved analytically \footnote{See details in the Supplemental Material}. The simulations include a Gaussian random disorder of correlation length $l_c$ and amplitude $V_0$, and take into account the finite aperture of the objective. We show in Fig.\ref{fig3} the results for the first-order correlation function (\ref{measured_g}) with parameters close to the experimental conditions and two different realizations of disorder for each condensate. We find that the mean-field model well captures the main features observed in the experiment  \footnote{A description beyond mean field is currently being developed, starting from the disorderless  case \cite{chiocchetta2013}.}. The decay of the correlations in the model, due to phase fluctuations, originates from the out-of-equilibrium nature of the condensate, under the combined effect of the random potential which breaks spatial inversion symmetry and the effect of the temporal average in Eq.(\ref{measured_g}).  The modulations are due to a shallow disorder of amplitude $\sim 2meV$, i.e. comparable with the polariton linewidth. In the model, the disorder sets also the width of the central peak, which in our measurements is of the order of the experimental resolution.  In this regime the interactions only have a marginal effect on the correlations. Simulations predict a peak broadening at stronger interactions or weaker disorder.

In conclusion, we have shown that polariton lasing in ZnO microwires yields the formation of a one-dimensional weakly-interacting condensate of quasi-excitonic nature. This peculiar feature is achieved thanks to weak excitonic interactions together with a large Rabi splitting, and a negligible excitonic disorder of the bulk semiconductor material. In spite of the tiny photonic fraction, the correlations length of this condensate is found to be in the ten micron range, i.e. comparable to that of a 50\% photonic polariton condensate. According to our mean-field analysis, the first-order coherence is dominated by disorder while interactions have a negligible effect. In this regime, the decay of correlations seems to be weakly affected by driven-dissipative induced noise \cite{chiocchetta2013} and quantum fluctuations \cite{Haldane1981}. Therefore, even heavier and more exciton-like - i.e. above $97\%$ - ZnO microwire polaritons should still be able to easily condense.

A.T. acknowledges financial support by the "Fondation nanoscience" RTRA contract No.FCSN 2008-10JE. E.D., F.M., A.M. and M.R. acknowledge financial support by the ERC grant "Handy-Q" No.258608. Enlightening discussions with Le Si Dang and M. Wouters are warmly acknowledged. We thank Prof. Zhanghai Chen for providing us with high quality samples.

\end{document}